\documentclass[sigconf]{acmart}

\AtBeginDocument{%
  \providecommand\BibTeX{{%
    \normalfont B\kern-0.5em{\scshape i\kern-0.25em b}\kern-0.8em\TeX}}}

\copyrightyear{2024}
\acmYear{2024}
\setcopyright{acmlicensed}
\acmConference[CHI '24]{Proceedings of the CHI Conference on Human Factors in Computing Systems}{May 11--16, 2024}{Honolulu, HI, USA}
\acmBooktitle{Proceedings of the CHI Conference on Human Factors in Computing Systems (CHI '24), May 11--16, 2024, Honolulu, HI, USA}
\acmDOI{10.1145/3613904.3642812}
\acmISBN{979-8-4007-0330-0/24/05}

\usepackage{enumitem}
\usepackage{makecell}

\begin{document}

\title[Understanding Nonlinear Collaboration between Human and AI Agents:\\
A Co-design Framework for Creative Design]{Understanding Nonlinear Collaboration between Human and AI Agents: A Co-design Framework for Creative Design}

\author{Jiayi Zhou}
\orcid{0000-0003-4669-4872}
\affiliation{%
  \institution{State Key Lab of CAD \& CG,\\ Lab of Art And Archaeology Image, Zhejiang University}
  \city{Hangzhou}
 \state{Zhejiang}
  \country{China}
}
\email{jiayizhou@zju.edu.cn}

\author{Renzhong Li} 
\orcid{0000-0002-6577-036X}
\author{Junxiu Tang}
\orcid{0000-0003-3594-926X}
\affiliation{%
  \institution{State Key Lab of CAD \& CG,\\ Zhejiang University}
  \city{Hangzhou}
  \state{Zhejiang}
  \country{China}
}
\email{renzhongli@zju.edu.cn}
\email{tangjunxiu@zju.edu.cn}

\author{Tan Tang}\authornote{corresponding author.}
\orcid{0000-0002-5260-3087}
\affiliation{%
  \institution{Lab of Art And Archaeology Image, Zhejiang University}
  \city{Hangzhou}
  \state{Zhejiang}
  \country{China}
}
\email{tangtan@zju.edu.cn}

\author{Haotian Li}
\orcid{0000-0001-9547-3449}
\affiliation{%
  \institution{The Hong Kong University of Science and Technology}
  \city{Hong Kong SAR}
  \country{China}
}
\email{haotian.li@connect.ust.hk}

\author{Weiwei Cui}
\orcid{0000-0003-0870-7628}
\affiliation{%
  \institution{Microsoft Research Asia}
  \city{Beijing}
  \country{China}
}
\email{weiweicu@microsoft.com}

\author{Yingcai Wu}
\orcid{0000-0002-1119-3237}
\affiliation{%
  \institution{State Key Lab of CAD \& CG,\\ Zhejiang University}
  \city{Hangzhou}
  \state{Zhejiang}
  \country{China}
}
\email{ycwu@zju.edu.cn}

%\thanks{*}

\renewcommand{\shortauthors}{Zhou, et al.}

\begin{abstract}
\label{sec:abstract}
Creative design is a nonlinear process where designers generate diverse ideas in the pursuit of an open-ended goal and converge towards consensus through iterative remixing.
In contrast, AI-powered design tools often employ a linear sequence of incremental and precise instructions to approximate design objectives.
Such operations violate customary creative design practices and thus hinder AI agents' ability to complete creative design tasks.
To explore better human-AI co-design tools, we first summarize human designers’ practices through a formative study with 12 design experts.
Taking graphic design as a representative scenario, we formulate a nonlinear human-AI co-design framework and develop a proof-of-concept prototype, OptiMuse. 
We evaluate OptiMuse and validate the nonlinear framework through a comparative study.
We notice a subconscious change in people's attitudes towards AI agents, shifting from perceiving them as mere executors to regarding them as opinionated colleagues. 
This shift effectively fostered the exploration and reflection processes of individual designers.
\end{abstract}




\begin{teaserfigure}
  \includegraphics[width=\textwidth]{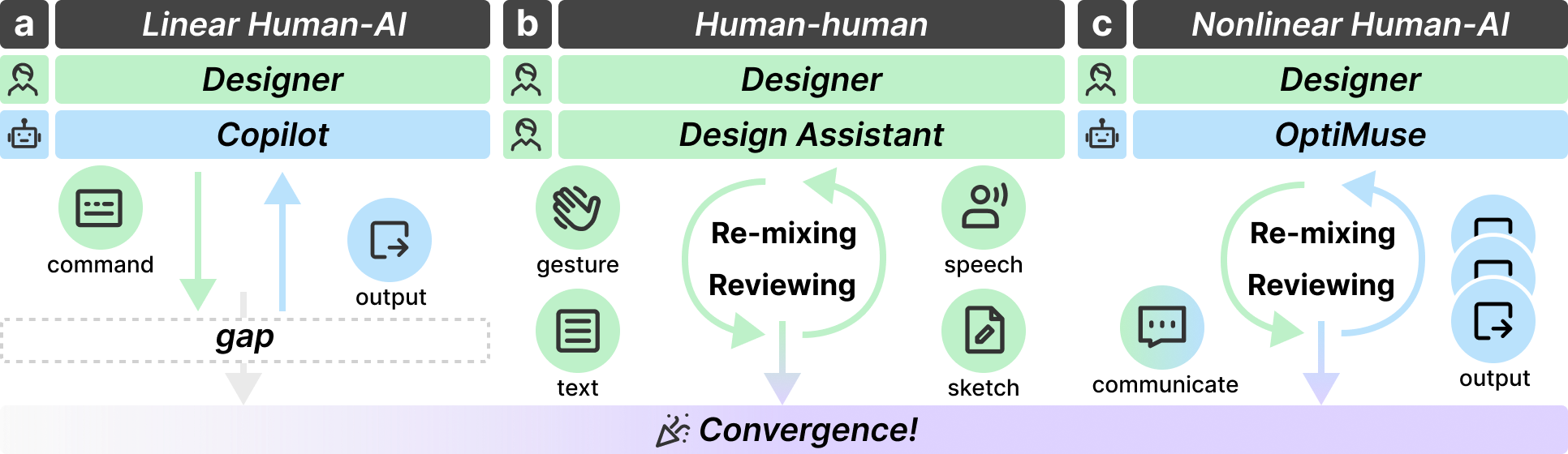}
  \caption{(a) pipeline of the traditional linear AI-driven tools; (b) a formative study involving human-human co-design; (c) a nonlinear human-AI co-design framework.}
  \Description{}
  \label{fig:teaser}
\end{teaserfigure}

%\received{20 February 2007}
%\received[revised]{12 March 2009}
%\received[accepted]{5 June 2009}

\maketitle

\section{Introduction}
\label{sec:intro}
Creative design unfolds with an open-ended goal, which cannot be achieved overnight.
Despite having an initial idea, designers can continuously generate fresh thoughts when they encounter inspiring stimuli~\cite{muller-wienbergen_leaving_2011}.
By fusing the strengths of different thoughts, designers strive to create a unified and cohesive output.
With the rapid advancement of AI technology, numerous generative design tools (e.g. Adobe Firefly~\cite{adobe}, Midjourney~\cite{midjourney}) have emerged to aid in the creative design process.
These tools can communicate with users and optimize design results according to textual commands. 
However, they may suffer from poor user experience (e.g. low success rates~\cite{buschek_nine_2021}) due to the ignorance of natural collaboration process in creative design~\cite{howard_describing_2008}.

According to previous studies~\cite{hatchuel_new_2003,pugh_total_1991,designcouncil,gero1990design,gero2004situated,howard_describing_2008}, the collaboration workflow in creative activities should be nonlinear.
Users progressively refine and align their open-ended requirements as well as remix different design ideas throughout the nonlinear collaboration process (Fig. ~\ref{fig:teaser}b). 
When using AI-assisted design tools, users often struggle to provide explicit and executable commands, resulting in abstract open-ended requirements~\cite{buschek_nine_2021}. 
Consequently, the quality of the results generated by AI is severely impacted, and it becomes challenging to inspire designers and assist them in remixing design ideas.
Compared to the nonlinear process which involves requirement alignment and option remixing, existing human-AI co-design practices mainly rely on command execution, resulting in a more linear workflow (Fig. ~\ref{fig:teaser}a).

In various related domains, such as requirement analysis~\cite{royce1987RA, ambreen2018RA, maguire2002RA, demirel2018RA, saraiya2005RA} and psychology~\cite{2019land, flower_cognitive_1981, reynante_framework_2021}, research has focused on studying collaboration patterns among humans and highlighting the importance of requirement alignment and remixing. 
However, the process of requirement alignment and remixing between humans and AI remains an under-explored problem.
To address this problem, we conducted a formative study consisting of two parts to learn from the human-human collaboration. 
Taking graphic design as a representative scenario, the first part involved pairing participants to collaboratively complete a creative design task.
The second part involved interviewing the participants to identify challenges in the collaboration process and gather their requirements for AI collaborators.
By observing the human-human collaboration process and analyzing the gathered feedback, we summarize five design requirements for nonlinear AI-assisted tools. 
These design requirements can be categorized into two main types: implementing nonlinear characteristics in human-AI collaboration (e.g., providing alternative design solutions for inspiration) and improving the shortcomings of human-human collaboration (e.g., achieving real-time visual outcomes).

Based on the formative study, we propose a novel framework (Fig. ~\ref{fig:teaser}c) that introduces the nonlinear human-AI collaboration for creative design tasks.
Our framework incorporates ten specific actions for AI agents centered around five areas (flexible communication, clarification, alternative solutions, operational strategies, and visual outcomes) corresponding to the design requirements.
To validate the effectiveness of the framework, we develop a proof-of-concept prototype (POC), OptiMuse, and evaluate it through a comparative experiment using the Wizard-of-Oz method.
To establish a baseline tool, we adopt the state-of-the-art tool, Office 365 Copilot for PowerPoint~\cite{copilot}, which follows the linear human-AI workflow.
Copilot provides users with an optimal output to their commands without explaining how it is inferred from the user’s instructions. 
We recruited 12 participants to complete creative design tasks using OptiMuse and Copilot.
The experiment results indicate that OptiMuse significantly improves users’ task completion rates.
Upon analyzing the interview data, we identify four expected roles of AI agents.
Our main contributions are as follows:
\begin{itemize}[nosep]
\setlength{\leftskip}{-12pt}
    \item We conducted a formative study to investigate the co-design process between human collaborators and formulated a set of design requirements for establishing human-AI co-design tools.
    \item We propose a human-AI co-design framework and develop a proof-of-concept prototype, OptiMuse.
    \item We conducted a comparative study to validate our framework and evaluate the usability of OptiMuse.
\end{itemize}

\section{Related Work}
\label{sec:relatedwork}
According to the well-established taxonomy~\cite{howard_describing_2008, wang_literature_2017}, we introduce relevant studies with regards to (1) the concept and practice of the creative design process, (2) the methodologies adopted by HCI researchers in developing creativity-support tools, and (3) the strategies utilized in human-human collaboration to foster creativity.

\subsection{Creative Design Process}
In the design process, creativity plays an irreplaceable role in promoting innovation~\cite{mumford_creativity_1988, yusuf2009creativity}.
The investigation of the creative process, as opposed to the conventional design approach, piques people's interest.
Some models~\cite{hatchuel_new_2003,pugh_total_1991, designcouncil,gero1990design,gero2004situated,howard_describing_2008} incorporate nonlinear design workflow that might give rise to creativity.
For instance, the ``C-K theory''~\cite{hatchuel_new_2003} portrays design as a nonlinear navigation within two spaces: the concept space (C) and the knowledge space (K). The ``concept space'' refers to a space of basic semantic meanings in the design process while the ``knowledge space'' means a logical space of designers' ``propositions''.
As this model depicts, the design process is dynamic, evolving between the two spaces.
Another design model is the Double Diamond~\cite{designcouncil} which, as a typical divergent-convergent model, includes exploring widely (divergent thinking) and taking focused action (convergent thinking). Moreover, the looped arrays in the model remind the designers that practices toward innovation are featured by repeated evaluation and iteration.
Gero's function, behavior, and structure (FBS) model~\cite{gero1990design} comprises eight fundamental design processes, with particular emphasis on the three reformulation processes. These processes underscore a dynamic design approach by emphasizing the significance of making changes in response to identified dissatisfaction.
Furthermore, the situated FBS framework~\cite{gero2004situated} sets a new foundation for intelligent design systems by leveraging the ``reformulation'' of agents in an open, dynamic world.
Combining the view of cognitive psychology, T. J. Howard et al.~\cite{howard_describing_2008} mapped analysis, generation, and evaluation as the consensus of the creative process into the FBS model.
The recognition and emphasis on the nonlinear nature of the creative process have been evident in various design models. However, the advancement of AI has rendered the need for a ``renaissance'' in a creative design framework with a focus on human-AI collaboration.

Certainly, while these nonlinear models may capture the essence of the creative design process, their high-level descriptions can not seamlessly translate to the intricacies of real-world design~\cite{howard_describing_2008}, especially for human-AI collaboration practices. 
Therefore, rather than strictly adhering to one of these nonlinear creative design models, we aim to glean insights directly from human-human design practices among design experts, enabling the formulation of models that better align with real-world scenarios.
\subsection{From Creativity Support Tool to Human-Computer Co-creativity}
How can computers actively engage in the creative design process to ignite and enhance creativity? The HCI community has investigated this issue based on the design lessons about developing creativity-support tools~\cite{wang_literature_2017,Althuizen_Creative_2016,clark_creative_2018,kulkarni2023word,oh_i_2018}. 
Tools that support creative tasks have been widely studied~\cite{wang_literature_2017,muller-wienbergen_leaving_2011}, including approaches supporting the creative process of finding problems, ideas, and solutions~\cite{wang_literature_2017}.
For example, based on the idea of ``machine-in-the-loop'' Clark et al.'s machine agent~\cite{clark_creative_2018} provides writing suggestions for human users to choose from, and users are always in control of the final results.
Oh et al.~\cite{oh_i_2018} developed DuetDraw, a creative drawing tool equipped with AI, to provide drawing ideas and directly draw on the canvas.
Moreover, by evaluating a set of AI-steering music design tools, Louie et al.~\cite{louie_novice-ai_2020} reported positive user feedback on human-AI collaboration.
Kulkarni et al.~\cite{kulkarni2023word} demonstrated that generative text-to-image models act as a helpful solution for novice users in design exploration and can promote human-AI collaboration.
Although design theories acknowledge the non-linear navigation between design stages~\cite{hatchuel_new_2003,pugh_total_1991, designcouncil,gero1990design,gero2004situated,howard_describing_2008}, these creativity-support tools often concentrate on examining one or two stages individually~\cite{wang_literature_2017} without a recognition of the nonlinear navigation between these stages.
This limitation impedes the transformation of computers from being viewed as mere tools for creative support to being perceived as inherently creative entities.

Recently, AI agents have become more integrated into the creative process, requiring new studies to promote natural human-AI collaboration.
Various HCI theories~\cite{lubart2005can,wang_literature_2017,muller-wienbergen_leaving_2011,negrete2014apprentice,kantosalo2016modes,guzdial2019interaction,rezwana2022understanding} have been conducted to provide guidelines for human-AI co-creativity.
The apprentice framework~\cite{negrete2014apprentice} analyzes creative processes as a collaborative effort between humans and computers, assigning them distinct roles and tasks.
It identifies four creativity pieces emerging during the creative design process and five roles a computer can play to produce them.
From a computational perspective, Kantosalo et al.~\cite{kantosalo2016modes}  defined \textit{alternating co-creativity} where human and AI equally take turns to finish creative tasks.
The process lifts strict requirements to \textit{complete creative agent} who must be capable of identification, generation, and evaluation.
Moreover, Guzdial et al.~\cite{guzdial2019interaction} presented a general framework for turn-based human-AI co-creativity by identifying components and design reflections.
A user study conducted by Rezwana et al.~\cite{rezwana2022understanding} demonstrates that participants perceive AI as more human-like—depicting qualities when facilitating AI-to-human communication.
However, these works mainly conceive of the nonlinear creative design process as a straightforward, turn-based interaction between users and computers.
By thoroughly examining human-human creativity practices, with a particular focus on nonlinearity, our study introduces a novel human-AI co-design framework that more accurately reflects the nonlinear navigation through the various stages of creative design.

\subsection{Collective Creativity}
As Maher~\cite{maher2012computational} stated, ``Creativity can ascribed to a computational agent, an individual person, collectives of people and agents and/or their interaction.''
The social nature of creativity, which highlights that much of our creative output is a result of interaction and communication with others~\cite{fischer2004social}, emphasizes the significance of collective creativity as a method that promotes innovation through collaborative efforts~\cite{Inakage2007collective}.
For example, the TickRok music Duets enable a collective exploration of design space to increase ideation creativity~\cite{O'Toole2023collaborative}.
Specifically, the ``cumulative'' evolution of creativity is achieved in the parallel interaction of multiple design versions.
Another example is the micro-task approach~\cite{Girotto2016collective}, a collective creativity strategy that enables synergistic collaboration by defining sub-tasks and assigning them to skilled and motivated communities.
Group collaboration that sparks creativity is always accompanied by ``remixing'', which is important and frequently observed in real-world creativity practices~\cite{Dynamicland, scratch, bruckman_community_1998}.
Many design tools allow designers to merge their human collaborators' ideas into their creative output iteratively~\cite{dasgupta_remixing_2016, bruckman_community_1998, marlow_rookie_2014}.
Also, there has been a surge in research interest in exploring methods to foster innovative remixing~\cite{tan2022conflict,Cheliotis2014remix,Oehlberg2015remix,yue2019improving}.
For instance, through an empirical study on the Thingiverse (a 3D printing community), Tan et al.~\cite{tan2022conflict} revealed the contradictory impact of knowledge endowment heterogeneity on users' quality and quantity of remixing.
However, there is a gap in understanding how AI can support people in remixing to enhance creativity in human-AI collaboration.
To fill this gap, we started with a typical scenario of creative design.
Creativity is a crucial aspect of graphic design, enabling designs to stand out from their competitors and capture viewers' attention~\cite{graphicdesign}.
Our study aims to enhance the understanding of collective creativity by proposing a collaboration framework for human-AI creative design and investigating how individual designers re-mix alternative designs when working on AI-assisted graphic designs.

\section{Study I: Understand the Nonlinear Process in Creative Design}
\label{sec:formativestudy}
\begin{figure*}
  \centering
  \includegraphics[width=1.0\textwidth]{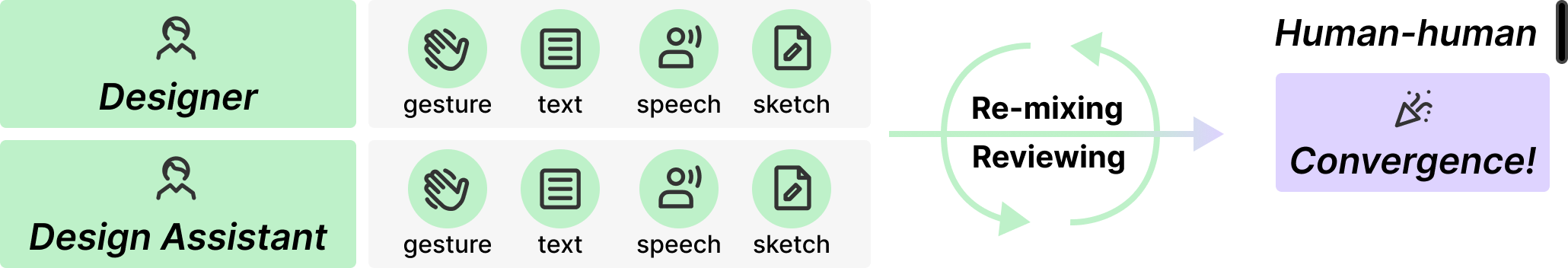}
  \caption{Creative design is a nonlinear process with an open-ended goal where almost no designer follows a predefined procedure. Designers search for multiple design options and jump from one sub-solution to another as they actively review, remix, and iterate them until achieving satisfactory results.}
  \label{fig:human-human creative design}
\end{figure*}
The aim of Study I is to establish guidelines for fostering better collaboration between human and AI agents during the creative design process.
Building upon the insights from existing study~\cite{wang_human-human_2020} that emphasize the integration of human-centered design philosophy into complex human activities, we first sought to gain a deeper understanding of human-human collaboration in creative design ~(Fig.~\ref{fig:human-human creative design}).
We observe users' co-design process when they face-to-face optimize graphic design.
Our expected results focus on:
\begin{enumerate}[nosep]
\setlength{\leftskip}{-12pt}
    \item How do designers express vague requirements and the related specific operations in natural language?
    \item How do vague requirements transform into specific operations?
\end{enumerate}
\vspace{-4pt}
\subsection{Preconsideration of Experiment Settings}
The creative process is a purpose-driven journey in which individuals construct a hierarchical network of goals to serve as their guiding framework\cite{flower_cognitive_1981}. Requirement analysis (RA), a thorough analysis of the goals, is important during the process because it helps ensure that the desired objectives and outcomes of a project are clearly understood and defined. While RA in the design process has not been extensively studied, it has been a topic of longstanding discussion in other domains, particularly in software development~\cite{ambreen2018RA, demirel2018RA, maguire2002RA, royce1987RA, saraiya2005RA}. Our formative study design incorporates the knowledge derived from RA.

To explore the possible task division between human and AI agent, we followed a formative study strategy presented by Pan et al.'s study~\cite{pan_human-computer_2023}.
We simultaneously observed the behaviors of designers and design assistants.
To obtain guidance from the collaborative process, it is important to develop scenarios that are conducive to collaboration.
We employed a role-playing strategy and set up the best collaboration scenario during the formative study.
The role-playing strategy also helped participants to recall relevant experiences.

A pilot study was conducted among three experienced design experts to help us determine the details of the procedure. They had professional experience in design for 3, 4, and 5 years, respectively.
Two experts and the experimenter gathered in person, assembling around a laptop.
They shared their screens with another expert who participated remotely through an online meeting platform. The experts engaged in discussions on design enhancements collectively without a predefined procedure.
Based on the natural co-design process shown in the pilot study, we established favorable collaborative conditions for the participants.
Because online meetings would hinder effective communication in co-design, the modification sessions were conducted in person, enabling multimodal interactions and real-time communication.
Additionally, we allocated time for content comprehension to ensure a fundamental understanding of the target design.
Afterward, designers brainstormed individually in preparation for the discussion.

\subsection{Experiment Design: A Formative Study}
To gain insights from experts’ practice, we conducted a human-human collaborative design study that was recorded for subsequent analysis with participants’ consent.
\vspace{-4pt}
\subsubsection{Participants}
    We recruited participants with a minimum of one year of design experience through a recruitment message posted on various online social media platforms.
    These individuals were further selected based on the quality of their design work.
    In total, our study involved 12 participants (7 women and 5 men). All of them had their designs showcased in prominent events such as public activities, design competitions, or art exhibitions with a substantial number of attendees exceeding hundreds.
    When recruiting participants, we collected 5 graphic designs and their modification requests from candidates.
    These designs involved topics of medicine, biology, literature, product design, game design, and so on.
    To ensure design diversity, we selected 3 out of 5 designs that exhibit variations in layout, target audience, and modification requests as formal study material.
\vspace{-4pt}
\subsubsection{Procedure}
The study started with an overview of the collaborative modification process.
We explained the modification process as a collective effort, where individuals collaboratively devise concrete modification strategies based on initially vague requirements.
To maintain coherence, the modification strategies needed to fall within the scope of single-page layouts.

In the modification session, two participants took part in the study simultaneously, one playing the role of \textit{designer} and the other of \textit{design assistant}.
The \textit{design assistant} was asked to modify the designs submitted by the \textit{designer}.
After that, they switched roles and repeated the modification session.
The 25-minute modification session was detailed as follows:

\begin{itemize}[nosep]
\setlength{\leftskip}{-12pt}
    \item \textbf{Content Understanding.}
    The \textit{designer} explains the content of the designs and demonstrates the requirements within 5 minutes while the \textit{design assistant} has the opportunity to interrupt and ask questions about design content.
    \item \textbf{Independent Modification Strategies Brainstorming.}
    After understanding the modification task, the \textit{design assistant} has 5 minutes to brainstorm potential modification strategies independently.
    The \textit{design assistant} is allowed to sketch or write on Apple Freeform as part of the brainstorming process.
    \item \textbf{Synchronization.}
    Then the \textit{designers} and \textit{design assistant} have 10 minutes to reach a consensus on the modification requirement and specific optimization strategies for each design.
    \item \textbf{Modificaiton Summarizing and Rating.}
    The \textit{design assistant} is asked to provide an oral summarization of the optimization strategies employed and describe how they reach a consensus.
    After that, the \textit{designer} provides feedback on the \textit{design assistant} using a 5-point rating scale, assessing the effectiveness of requirement analysis and fulfillment.
\end{itemize}

We then proceeded with a semi-structured interview to gain deeper insight into the scenario and further explore users' engagement in the future utilization of AI agents in design modification tools. We asked each participant the following questions: (1) \textit{What are the challenges you encountered when collaborating with others in design modification and during this experiment? What is your role during that process?} (2) \textit{What are the advantages and disadvantages of collaborative design compared to independent design?} (3) \textit{What assistance do you expect from AI when modifying the design? Are there any tasks or aspects that you prefer to handle independently? If so, what are the reasons behind your preference?} Several additional questions were tailored based on the participant's performance during the modification session.

\subsection{Expert Feedback and Discussion}
Based on collected RA data in modification sessions and interview recordings, we summarize experts' feedback on the characteristics of requirements, as well as the pros and cons of the nonlinear human-human co-design process. Subsequently, we outline five design requirements (Table~\ref{tab:DR}) for Human-AI Co-design.
\subsubsection{The Inherent Nonlinearity of Requirements.}
We employed two approaches to summarize the characteristics of requirements: (1) We analyzed the evolution of design requirements during the \textit{Modification Session}; (2) We gathered feedback related to requirements analysis through \textit{Semi-structured Interviews}.
The process of requirement evolution is chaotic.
When modifying designs, requirements evolve, and users gain more insight into what they need in trial and error.  The inherent nonlinearity of requirements are reflected as: \textbf{requirement ambiguity}, \textbf{requirement changes}, and \textbf{requirement conflicts}.
Participants who played the role of designers in the first session were labeled as A, whereas the partner participants were labeled as B.
Specifically, $A_{ij}$ refers to the designers in the $i$-th group with regards to the $j$-th design. 
For further details, the relevant words of experts are provided in Appendix~\ref{app:feedback}.
\begin{itemize}[nosep]
\setlength{\leftskip}{-12pt}
    \item {\bf Requirement Ambiguity.} The initial version of design requirements was often underdeveloped and prone to ambiguity.
    In $A_{12}$ and $B_{23}$, designers voiced dissatisfaction without providing any optimization intentions.
    In other cases, optimization intentions were expressed in a general manner.
    $A_{32}$ and $A_{33}$ described the visual effect of a picture as a vague sensation rather than a visible outcome.
    \textit{``Identifying''} ($A_{11}, A_{23}, A_{41}, B_{61}, B_{63}$) and \textit{``improving''} ($A_{42}, A_{53}, A_{63}, B_{12}, B_{31}, B_{42}$) pages are the most frequent ambiguous descriptions.
    Compounding the issue, some participants ($A_{11}, A_{23}, B_{61}, B_{12}, B_{41}, B_{42}$) failed to clarify specific parts of the designs to be optimized.

    \item {\bf Requirement Changes.}
    While a few requirements ($B_{32}$, $B_{52}$) were fulfilled in a single iteration, the majority of designs (34 out of 36) underwent trial-and-error iterations.
    Some shifts in design requirements are generated when correcting inaccurate descriptions.
    Some requirements undergo iterations, while new requirements emerge in designers' minds, as a result of interpersonal dynamics within the design team. 
    In communication, cooperators make compromises to harmonize their perspectives and reach a consensus on design requirements.
    \item {\bf Requirement Conflict.}
    Requirements are constantly subject to negotiation and often clash with one another.
    We identify three types of conflicts:
    (i) \textit{Conflicts between asynchronous requirements}, which arise naturally as designers change their minds, normally designers focus on current ones;
    (ii) \textit{Conflicts between individuals}, which frequently happen because different people see things from different perspectives;
    and (iii) \textit{Conflicts within the design itself}, which involve factors such as functionality, aesthetics, budget, time constraints, user preferences, technical feasibility, and ethical considerations.
    Striking a balance between conflicting requirements involves exploring alternative design options, presenting selective outcomes for feedback, and iterating on the design based on negotiation.
    It is crucial to prioritize the core requirement while accommodating as many perspectives as possible.
\end{itemize}
\subsubsection{Design Requirements Inspired by Human-Human Co-design.}

The co-design process is an imaginative voyage of discovery. 
According to the interview, the presence of another designer brings forth a multitude of distinctions and divergences in many aspects, which brings about both pros (+) and cons (-) of the nonlinear human-human co-design process.
\begin{itemize}[nosep]
\setlength{\leftskip}{-12pt}
    \item[+]{\bf Inspire to design in alternative ways.}
    In practice, every designer's comprehension of design is restricted.
    Despite their extensive knowledge and skills, designers' approaches to modification are confined by their understanding of the requirements.
    The inclusion of an extra viewpoint from a collaborator can stimulate creativity. Our study demonstrates that unanticipated inspiration from others enables achieving design goals through alternative methods.
    \item[+]{\bf Accelerate design progress in timely communication.} 
    When designers refine designs independently, they sometimes are stuck in minor details.
    In contrast, when collaborating, the need for prompt communication necessitates swift responses.
    Consequently, ideas are rapidly iterated until both designers and design assistants reach a consensus.
    \item[+]{\bf Expose and fix misunderstandings through Q\&A.} 
    Frequent sync allows people to detect and react to misunderstandings timely.
    During communication with their design assistants, designers are actively thinking about how to convey the primary concept of designs.
    By observing their partners' reactions, they become aware of potential misunderstandings on the designs.
    In turn, design assistants gain a proper understanding as misconceptions are rectified during the conversation.
    \item[-]{\bf Lack real-time outcomes.}
    Despite all the pros, designers report a lack of real-time outcomes in human-human collaboration. Due to the time-consuming manual modification, participants can only imagine a visual outcome according to words and sketches, which takes a toll on accuracy.
\end{itemize}

To strengthen the pros and address the cons, we further summarize five design requirements (table ~\ref{tab:DR}) for a human-centered nonlinear framework to support better human-AI co-design.

\begin{table*}[t]
\centering
\caption{Design requirements for Human-AI Co-design systems.}
\begin{tabular}{p{.81944cm} p{16.1111cm}}
\hline
\textbf{ID}  & \multicolumn{1}{c}{\textbf{Design Requirements}} \\
\hline
\multicolumn{2}{p{16.8506cm}}{(i) Strengthen the pros of the nonlinear human-human co-design process, with special attention to the nonlinear characteristics of requirements.} \\
\hline
\textbf{DR1} & Allow flexible communication to foster in-depth discussions concerning design requirements. \\
\textbf{DR2} & Encourage users to double-check possible misunderstandings. When misunderstandings happen, provide alternative explanations or allow users to provide clarification. \\
\textbf{DR3} & Offer alternative solutions as sources of inspiration. \\
\hline
\multicolumn{2}{p{16.8506cm}}{(ii) Address the cons of the human-human co-design process.} \\
\hline
\textbf{DR4} & Promote operational strategies according to the modification intentions. \\
\textbf{DR5} & Offer visual outcomes when promoting operational strategies. \\
\hline
\end{tabular}
\label{tab:DR}
\end{table*}

\section{A Human-AI Co-design Framework for Creative Design}
\label{sec:framework}
To fulfill the design requirements, we formulate a nonlinear human-AI co-design framework that incorporates specific actions for AI agents.
The framework is implemented in a POC called OptiMuse, where we focus on user interface design. Two primary components of the nonlinear framework are detailed in the subsequent sections.

\subsection{Expected Actions for AI Agents}
\label{sec:Ax}
Drawing from the formative study (Sec.~\ref{sec:formativestudy}), we summarize 10 expected actions designers hold for collaborators.  
Realizing these expectations can significantly enhance the co-design experience, a facet not yet fully actualized in current human-AI collaborations~\cite{lubart2005can,wang_literature_2017,muller-wienbergen_leaving_2011,negrete2014apprentice,kantosalo2016modes,guzdial2019interaction,rezwana2022understanding}. 
These expected actions, which are denoted as A1-A10, are primarily centered around DR 1-5: flexible communication, clarification, alternative solutions, operational strategies, and visual outcomes.

Concerning the first two design requirements, designers are more inclined to confidently express ambiguous descriptions if they have opportunities to provide explanations and rectify misunderstandings.
To facilitate the use of ambiguous commands, a communication process before AI agents generate outcomes is essential~\cite{shi2023Understanding}. 
To be specific, we emphasize the necessity of the following AI agent actions within the \textbf{communication process}:
\begin{itemize}[nosep]
\setlength{\leftskip}{-8pt}
    \item[\textbf{A1}]Guide and support users in refining ambiguous descriptions. (Corresponding to DR1, DR2)
    \item[\textbf{A2}]Provide reminders to users when requirements change or conflicts arise. (DR1)
    \item[\textbf{A3}]Demonstrate alternative explanations and descriptions of users' commands. (DR2)
\end{itemize}

\begin{figure*}[t]
  \centering
  \includegraphics[width=1.0\textwidth]{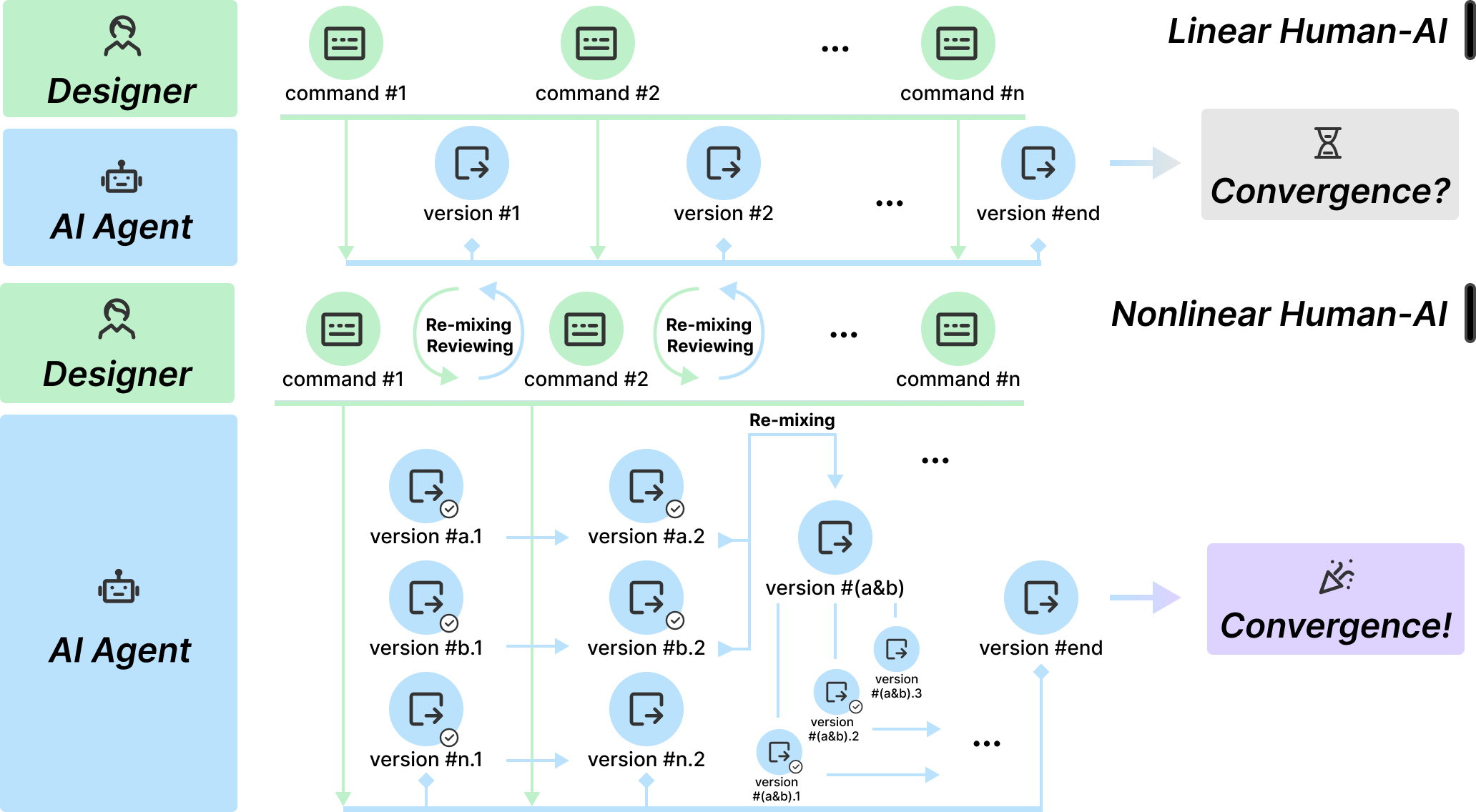}
  \caption{In the traditional linear human-AI collaboration, AI agents execute a series of gradual and precise commands to achieve final design outcomes. We propose a nonlinear human-AI co-design framework characterized by facilitating a communication process before producing visual results and accommodating versatile actions for multiple choices.}
  \label{fig:pipeline}
\end{figure*}

Regarding the third design requirement, design assistants present multiple selections to meet requests, based on design knowledge or existing examples. 
These selections were presented through a combination of textual descriptions, verbal explanations, sketches, supported gestures, and comprehensive visual results~\cite{shi2023Understanding} as design choices.
Designers derive inspiration when navigating among these design choices.
Navigating, summarizing, and discarding among the design choices are central to the nonlinear process of collective creativity ~\cite{Inakage2007collective,tan2022conflict,Cheliotis2014remix,Oehlberg2015remix,yue2019improving}.
The following actions were frequently observed in the formative study and, therefore, documented as required actions of AI agents in generating \textbf{multiple selections}:
\begin{itemize}[nosep]
\setlength{\leftskip}{-8pt}
    \item[\textbf{A4}]Revise users' requirements and offer different or additional choices. (DR3)
    \item[\textbf{A5}]Review choices to identify difference or superior options. (DR3)
    \item[\textbf{A6}]Modify the optimal choice for enhanced results. (DR3)
    \item[\textbf{A7}]Make revisions to multiple options and conduct subsequent comparisons. (DR3)
    \item[\textbf{A8}]Merge elements from different options. (DR3)
\end{itemize}
In alignment with the fourth and fifth design requirements addressing the cons of human-human collaboration, two required actions of AI agents are notable for \textbf{outputs}:
\begin{itemize}[nosep]
\setlength{\leftskip}{-8pt}
    \item[\textbf{A9}]Clarify operational strategies. (DR4)
\end{itemize}
\begin{itemize}[nosep]
\setlength{\leftskip}{-3pt}
    \item[\textbf{A10}]Preview the visual effect of operational strategies. (DR5)
\end{itemize}
These ten proposed AI actions are specifically crafted to address Design Requirements 1-5, striving to provide designers with a nonlinear experience that more closely resembles the flexibility and inspiration in the human-human collaboration process (Fig.~\ref{fig:pipeline}).

\begin{figure*}
  \centering
  \includegraphics[width=1.0\textwidth]{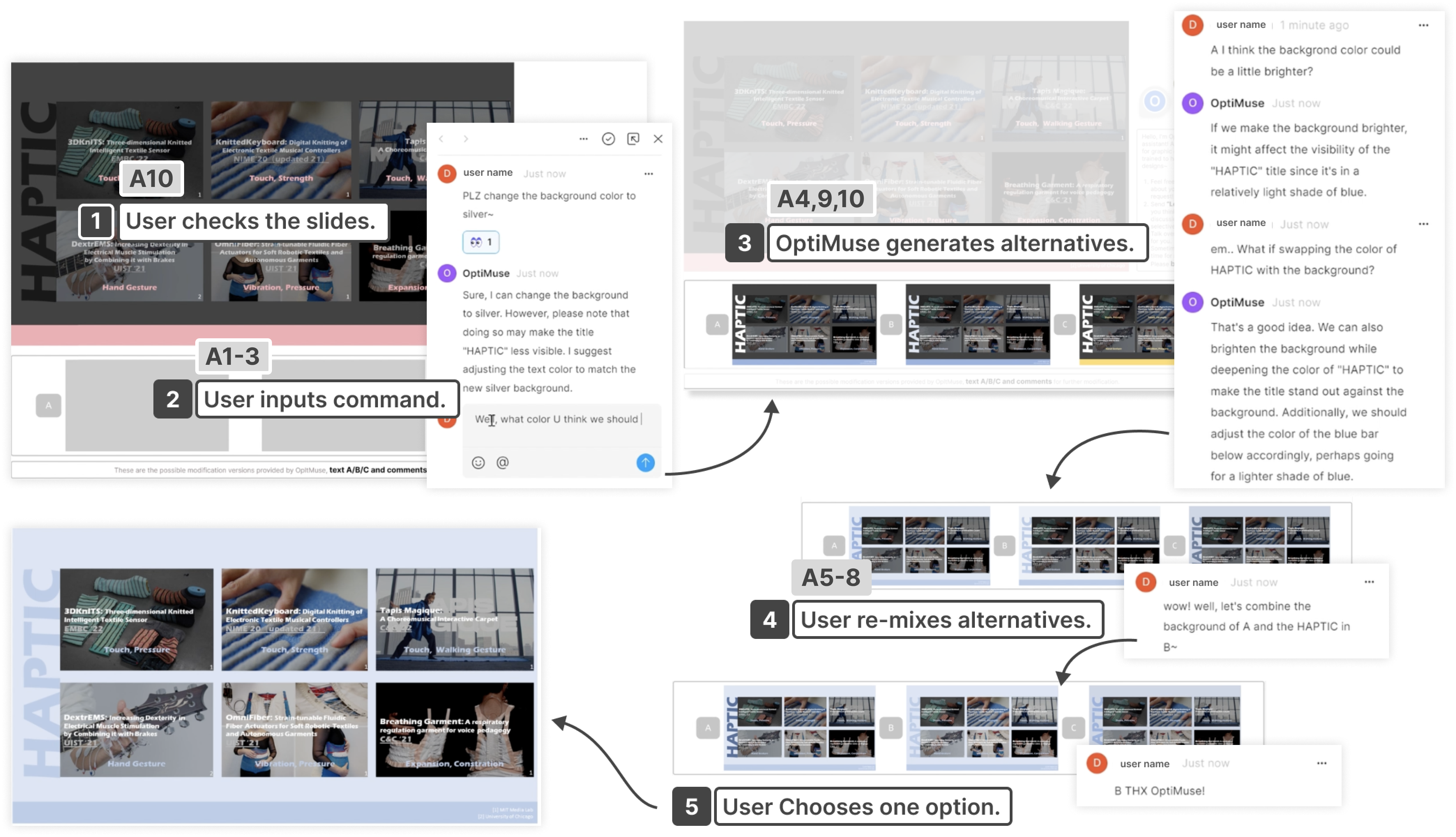}
  \caption{User interactions in OptiMuse follow these steps: (1) users review slides on the preview window; (2) users input commands and engage in rule-based conversation with OptiMuse; (3) OptiMuse generates alternatives as selective choices upon reaching convergence with users; (4) users remix choices through rule-based conversation; (5) users are satisfied with the choices and choose one option after several rounds of steps 1 to 4 (Notes: A1-10 refer to the AI agent actions defined in Sec.~\ref{sec:Ax}).}
  \label{fig:interaction}
\end{figure*}
\subsection{OptiMuse: a Proof-of-concept Prototype}
To facilitate effective interaction (Fig.~\ref{fig:interaction}) between designers and AI, we have developed the user interface of OptiMuse  (Fig.~\ref{fig:woz_OptiMuse}). 
This user interface comprises three primary components: a preview window, a dialogue box, and selective choices. 

The \textbf{preview window} displays the latest output of OptiMuse, offering users a dynamic visual representation of the design (Corresponding to \textbf{A10}) as it evolves through interaction.

Regarding the \textbf{dialogue box}, this is where designers communicate with the AI agent. 
Study I identified three dominant forms of flexible communication: \textit{text}, \textit{speech}, and \textit{draft}. 
However, due to the higher complexity in speech and draft interactions, text-based communication is more common in popular AI-driven tools~\cite{copilot,midjourney}.
Designers are accustomed to engaging in prompt-based communication with AI agents and expecting visual outputs due to the prevalence of prompt-based text-to-image models (e.g. Imagen~\cite{google2022imagen}, Parti~\cite{google2022parti}, and DALL-E 2~\cite{openai2022dalle2, openai2022dalleblog}). 
Consequently, we opted for text-based dialogue as the basic form of communication process.
To allow the required actions of AI agents (\textbf{A1-3}) for the communication process, we designed specific \textbf{dialogue regulations} in OptiMuse (see Appendix~\ref{app:wizard} for detailed rules for AI agents) :
designers engage in conversations with OptiMuse to align on optional strategies before they generate corresponding visual outcomes. 
They are mandated to transmit the signal \textit{`` Let's Have a try! ''} whenever they wish, indicating to OptiMuse to generate visual outcomes. 
Subsequently, they can issue commands and make reference to one or multiple options provided by it to continue their interactions based on the available choices. 
If they refer to only one option, the current outcome will be updated to reflect their preference. 
In addition, if they are dissatisfied with any of the provided choices, they have the option to signal \textit{`` NO ''} along with a new command to continue the discussion based on the existing outcome. 
Once they are content with the current outcome, they are required to transmit \textit{`` THX OptiMuse! ''} to signify the conclusion of the design process.

As for the \textbf{selective choices}, OptiMuse presents the selectable visual outcomes (\textbf{A9,10}) provided by AI agents. This design preserves the nonlinear nature of the creative design process while addressing the limitations of traditional human-human collaboration.
The selections encompass different design parameter options, interpretations of requirements, modification strategies, and combinations thereof (\textbf{A4}). To balance attention and choice overload~\cite{chernev2015choice}, we limit to three selective options to highlight the features of multiple choices. OptiMuse generates three distinct visual outcomes for each selection, with additional choices potentially emerging during dialogue. Users can request textual descriptions of these choices for review (\textbf{A5}).
The content of the selective choices is determined by the current outcome, the dialogue that occurs after the last \textit{`` Let's have a try! ''}, and OptiMuse's design knowledge.
To foster creativity, OptiMuse not only executes users' commands but also has the capacity to modify other aspects of the design, as long as these modifications align with design principles and do not contradict users' requirements.
If required by users, OptiMuse can perform further operations of reviewing (\textbf{A5}), modifying (\textbf{A6}), comparing (\textbf{A7}), and remixing (\textbf{A8}) the selective choices.
Together, the user interface design of OptiMuse allows for natural, nonlinear interactions between human and AI agents (Fig.~\ref{fig:interaction}) to perform tasks in creative design.

\section{Study II: Evaluation of OptiMuse}
\label{sec:userevaluation}
The purpose of Study II is to evaluate the usability of OptiMuse and validate the effectiveness of the novel human-AI collaboration framework.
We examine users' design modification processes when performing specific tasks on given graphical designs.
\subsection{Preconsideration of Experiment Settings}
\subsubsection{Choice of Modification Tasks and Design Materials}
We use Wizard of Oz (WoZ) experiments for Study ~II.
The reasons are threefold.
First, Copolit was not available since it had not been released to the public when we conducted Study II.
Second, our target is to assess the two workflows of co-design collaboration.
WoZ experiment allows us to manage the agent's output as expected under specific conditions.
Third, in WoZ environments, we can control the precise timing of each task.
It is important to introduce supplementary constraints through scenarios in WoZ experiments~\cite{Norman1991wizard, dahlback1993wizard, green2004wizard, Riek2012wizardscenario}.
To do so, we analyzed the graphical designs, corresponding requirements, and requirement-operational strategies chain collected from Study I. First, we categorized the modification tasks (requirements) based on levels of ambiguity and feasibility.
Then, we selected two specific tasks in graphical design: modifying the color scheme and adjusting the text layout because
(i) These tasks possess a certain degree of ambiguity, requiring exploration and investigation;
(ii) They limit the scope of modification possibilities, which makes it feasible to predict the requirement analysis process;
(iii) They frequently existed in Study I, suggesting their natural occurrence in the co-design practice.
To avoid unequal familiarity with designs, users were required to use different co-design tools to complete the same task on different designs.
Upon selecting two tasks, we carefully prepared two visually appealing graphical designs that were suitable for both tasks. 

\subsubsection{Wizard}
To ensure a valid WoZ simulation, we adhered to the guidance provided by Fraser and Gilbert~\cite{Norman1991wizard} in selecting, regulating, and training the wizard.
A qualified co-design assistant in study II requires a strong design background.
Thus, we enlisted an experienced design professional with extensive experience as a design leader in the university students' union for 2 years. This individual possesses a strong background in effectively communicating design requirements with supervisors and assisting fellow colleagues in accomplishing modification tasks.
Second, to ensure A1-10 while making the output from the wizard resemble that of a computer as far as possible regarding timing and consistency, we established specific guidelines for the assistant to minimize randomness during the communication and recommendation process.
Furthermore, the assistant was instructed to adopt certain behavior patterns commonly associated with AI agents. This was done to create an authentic interaction experience for participants, enhancing the perception that they were interacting with an AI agent.
Third, to ensure consistency in understanding, communication skills, and response time, we recruited only one wizard user II Additional details of guidelines for the wizard are provided in Appendix~\ref{app:wizard}.

\begin{figure}[t]
  \centering
  \includegraphics[width=0.478\textwidth]{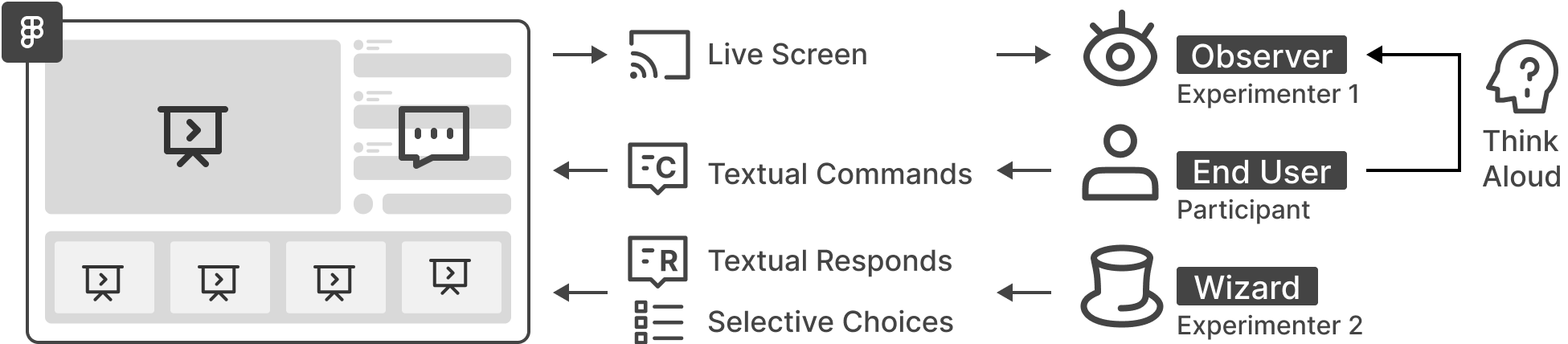}
  \caption{An overview of the scenario of Study II.}
  \label{fig:woz_workflow}
\end{figure}
\subsection{Study Design: A Wizard-of-Oz Study}
\subsubsection{Possible Modification Outcomes}
To enhance the design assistant's response time, we anticipated design requests and outcomes in advance. Three design students and three frequent AI-driven design tool users brainstormed individually, listing as many modification requests as possible in text and draft. Then, they gathered together in online meetings to discuss and organize potential modification requests. Finally, they prepared the corresponding design outcomes according to the modification requests.

\subsubsection{Participants}
We recruited 12 participants (4 males and 8 females).
They have design experience of 1-6 years, with 3.47 years on average. All of them had human-human co-design experience in modifying graphic designs before the experiment. The participants were randomly allocated into four groups of equal size.

\subsubsection{Baseline}
\begin{figure}[t]
  \centering
  \includegraphics[width=0.47\textwidth]{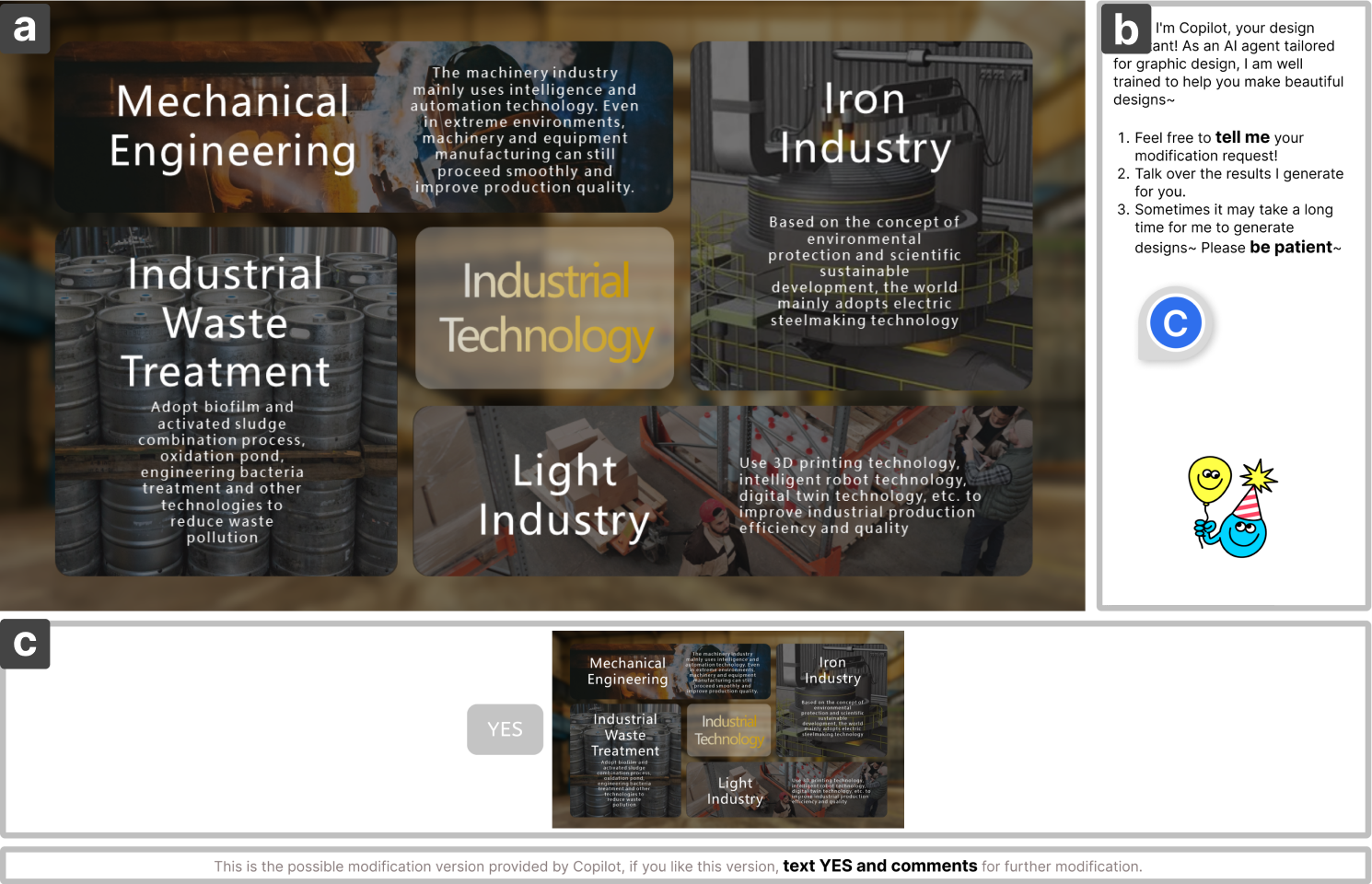}
  \caption{Overview of the baseline in Study II, Copilot: (a) The preview window shows the current outcome; (b) Users can type commands to Copilot; (c) Copilot produces a single visual outcome in response to each command.}
  \label{fig:Woz_Copilot} 
\end{figure}

\begin{figure}
  \centering
  \includegraphics[width=0.47\textwidth]{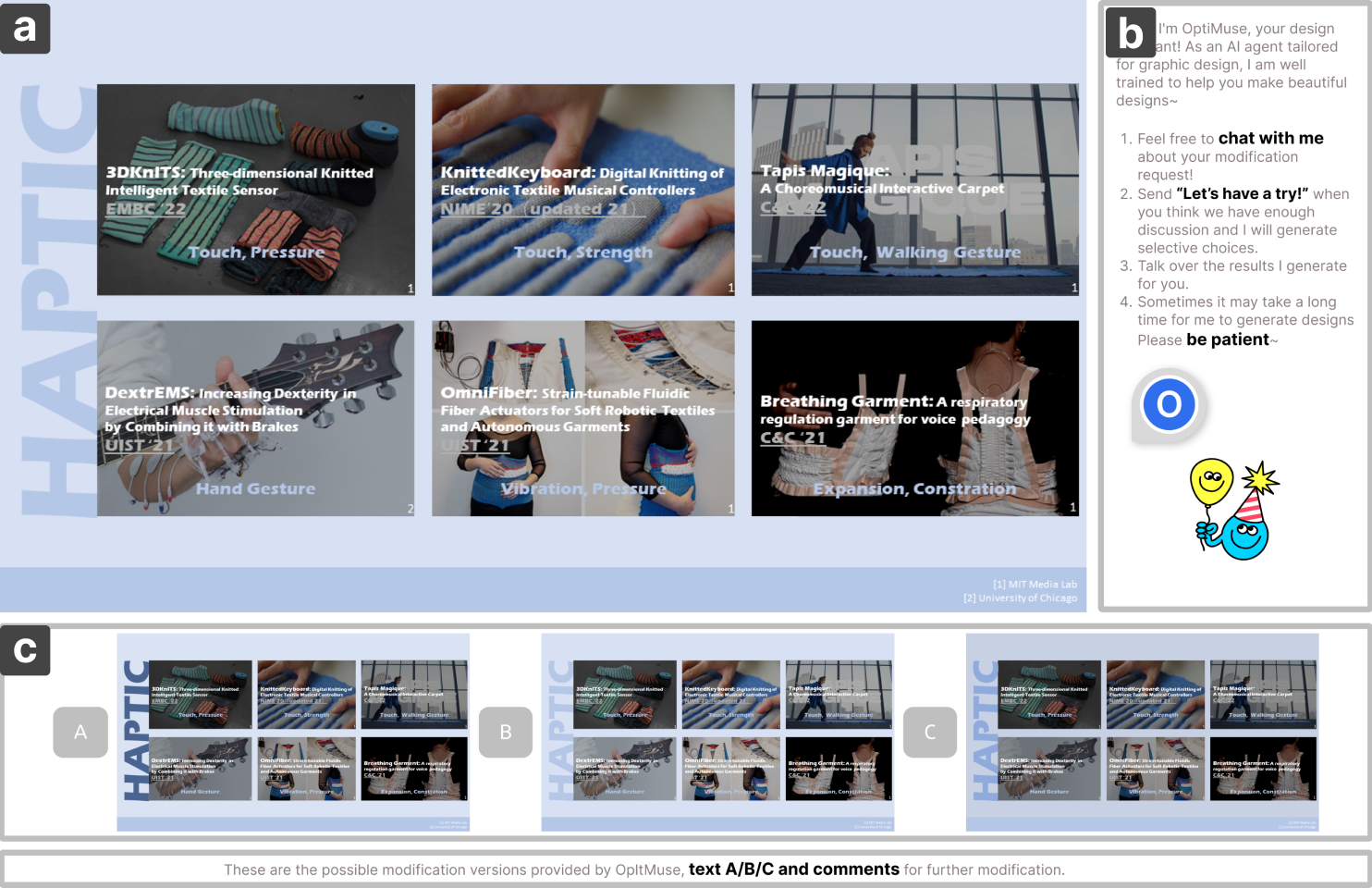}
  \caption{Overview of the OptiMuse : (a) The preview window shows the current outcome; (b) Users can communicate with OptiMuse through dialogues; (c) Copilot produces multiple visual outcomes when users type ``\textit{Let's have a try!}''.}
  \label{fig:woz_OptiMuse}
\end{figure}

We conducted a comparison between OptiMuse (Fig.~\ref{fig:woz_OptiMuse}) and Copilot (Fig.~\ref{fig:Woz_Copilot}), which is a prototype lacking the dialogue function and generating a single outcome each time.
Much like other widely used AI-powered design tools, including the publicly released demos of Microsoft PowerPoint Copilot\cite{copilot}, our Copilot operates by executing commands.
~

\subsubsection{Procedure}
The study contained two parts for each participant (Fig.~\ref{fig:woz_workflow}).
In the first part, each group of participants completed two design modification tasks.
The tasks were described as\textit{``Color Optimization: You can modify the colors of all elements in this design. The elements include background, text, and pictures. The color parameters include Hue, Saturation, Lightness, etc.''} and \textit{``Text Formatting: You can modify the formatting of all fonts in this design, including position, size, alignment, bold, italics, etc.''}.
In the second part, the participants utilized a different tool to complete the same two tasks. 
The order of tools and graphical designs used for each group is displayed in Table ~\ref{tab:task_order}.
Participants finished 48 (12 users * 2 tools * 2 tasks) tasks in total.
Apart from the Wizard, Study II involved an observer to observe the users' screen silently, listen to the users' thinking aloud, and mark the time of the interesting conversations.

Upon completing the four tasks, the participants were requested to fill out two System Usability Scale (SUS) questionnaires~\cite{SUS} to assess the usability of two tools (OptiMuse and Copilot). At the end of the experiment, we conducted a follow-up interview
and asked each participant the following questions:
(1) \textit{Reviewing this experiment, describe the challenges you encountered while using Copilot and OptiMuse.} (2) \textit{What are the advantages and disadvantages of using OptiMuse compared to Copilot in co-design tasks? Please clarify the differences first. What are their performances in handling requirements ambiguity, conflicts, and changes? How do selective choices make a difference? How does Q\&A make a difference?} (3) \textit{What specific assistance would you expect from AI agents like Copilot and OptiMuse in co-design? Are there any tasks or aspects that you prefer to handle independently, considering that they might exceed the capabilities of AI assistants? If so, what are the reasons behind your preference?} (4) \textit{Which collaboration mode is more similar to human behavior? What are the differences and pros and cons between the two collaboration modes and human-human collaboration?}
The study lasted 90 minutes. No participant expressed doubt regarding the presence of a wizard behind AI agents.
\begin{table}[h]
  \centering
  \caption{Task order for different groups.}
  \setlength{\tabcolsep}{4pt}
  \begin{tabular}{ccccccc}
    \hline
    Group & tool & design & Task & tool & design & Task \\
    \hline
    1 & OptiMuse & 1 & 1, 2 & Copilot & 2 & 1, 2 \\
    2 & OptiMuse & 2 & 1, 2 & Copilot & 1 & 1, 2 \\
    3 & Copilot & 1 & 1, 2 & OptiMuse & 2 & 1, 2 \\
    4 & Copilot & 2 & 1, 2 & OptiMuse & 1 & 1, 2 \\
    \hline
  \end{tabular}
  \label{tab:task_order}
\end{table}

\subsection{Results}
This section starts by presenting quantitative findings including SUS scores, time, dialogue turns, and task failure. Additionally, we discuss the disparities in requirements between OptiMuse and Copilot.
We then illustrate how the gathered data and interviews reveal people’s divergent attitudes toward OptiMuse and analyze their interaction patterns.
    
\subsubsection{Quantitative findings}
The average SUS score for OptiMuse is 70.21 (Good), while for Copilot, it is 64.17 (Poor). This indicates an improvement in the usability of our nonlinear human-AI collaboration as compared to the conventional linear approach.

\begin{figure}[ht]
    \centering
    \includegraphics[width=1\linewidth]{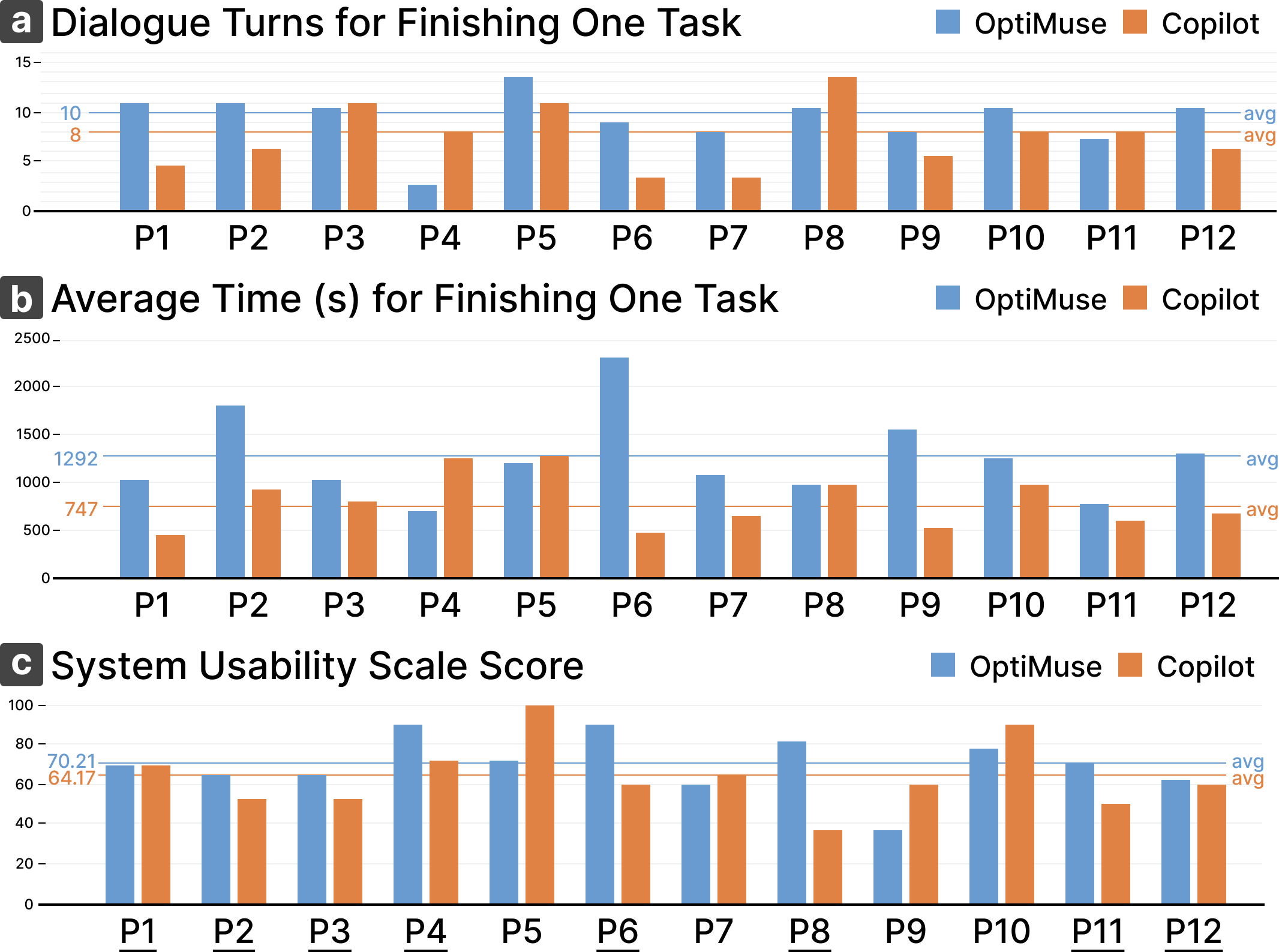}
    \caption{(a) Dialogue turns and (b) average time in finishing one task for each participant using OptiMuse and Copilot. (c) The System Usability Scale (SUS) score of each participant shows a divergent attitude toward OptiMuse (Notes: Underlined participants think OptiMuse is better than Copilot in terms of SUS score).}
    \label{fig:Quantitative_findings}
\end{figure}

We adhere to the instructions provided in the reference\cite{results} for presenting our quantitative findings in time and dialogue turns. Dialogue turns were computed by counting how many commands each participant sent to the AI agent. $P_{4}$ is an outlier who happened to be quite fond of the design provided by OptiMuse, so the process concluded quickly. After removing $P_{4}$, we obtained the following results: (i) Participants spent an average time of 1292 seconds and 747 seconds using OptiMuse and Copilot to complete one task, respectively. 
A Wilcoxon Signed-rank test shows that there is a significant effect of tools (W = 2, Z = -2.76, p < 0.05, r = -0.9). (ii) The average turns of OptiMuse and Copilot were 10 and 8 rounds. A Wilcoxon Signed-rank test shows that there is a significant effect of tools (W = 10, Z = -2.04, p < 0.05, r = -0.6).
$P_{1,5,7,8}$ gave up on tasks when using Copilot as they lost confidence in AI's ability to understand their design intentions and found it annoying in wording after an average of 1031 (365, 1803, 659, 1297) seconds and 8 (3, 10, 5, 15) turns. In conclusion, time, dialogue turn, and failure rate show that participants had greater tolerance toward OptiMuse in comparison to Copilot (Fig.~\ref{fig:Quantitative_findings}).

\subsubsection{Divergent attitudes towards OptiMuse and the interaction patterns}
Both SUS results (Fig.~\ref{fig:Quantitative_findings}) and interviews show divergent attitudes towards OptiMuse. While some participants were inspired by OptiMuse's textual feedback and selective choices, others were anxious and reluctant to engage in conversation.

\textbf{Positive feedback: fostering creativity and emulating human interaction.}
Individuals who praised OptiMuse were pleasantly surprised by the way it reshaped their initial design objectives. As $P_{10}$ mentioned: \textit{`` I was a little surprised when OptiMuse pointed out the problem of my requirements and offered alternatives. But it seems okay to give it a shot.''} They became receptive to OptiMuse's feedback, maintaining an open-minded perspective. They came to trust OptiMuse due to its logical and actionable feedback, as well as some gratifying selective choices. Despite the presence of conflicts between texts and visual output or misunderstandings, participants were enthusiastic about engaging with OptiMuse and allowed it to undergo trial and error. \textit{`` Indeed it appeared that OptiMuse did not do exactly what he said, but at the very least, the textual feedback provided me with a fundamental idea of what I could expect. ~''} We noticed some interesting \textbf{interaction patterns} in these users:
\begin{itemize}[nosep]
\setlength{\leftskip}{-14pt}
    \item\textbf{Asking questions:} Users frequently inquired about OptiMuse's opinion in the dialogue.
    When reluctant to articulate potential operations, they asked: \textit{`` What steps can I take in this scenario? Any suggestions?''}
    When seeking feedback and validation, their asked: \textit{`` How do you view my proposals?''}
    When faced with the complexity of discerning disparities among various options, they asked: \textit{`` What contrasts exist among these alternatives?''}
    \item\textbf{Remixing alternative choices:} Participants got inspiration and new design ideas by remixing design choices. Actions of remixing, as a commonly acknowledged approach to fostering creativity, were also frequently spotted in our participants.
    There were mainly two means to achieve remixing among several choices.
    First, remixing was observed through textual means. Our research yielded findings consistent with those of Kulkarni et al.~\cite{kulkarni2023word}. Textual content lends itself to a more straightforward form of remixing. During active discussions of potential operations, users amalgamate various ideas by rephrasing them into a new description. A simple action of copy worked well.
    Second, remixing involves adding commands to multiple choices, aligning with designers' inclination to draw insights from various sources. When they found satisfaction in specific aspects of different choices, they endeavored to identify those elements and merge them into a more improved solution. 
    However, despite describing the custom of engaging in such interactions based on past design practices and emphasizing their importance due to prior experiences with co-design AI tools, $P_{5}$ didn't actually carry out remixing of multiple selections. This was because he believed it wasn't possible (despite our instruction), largely because popular AI tools didn't support such actions.
\end{itemize}

\textbf{Negative feedback: too much AI and time waste.}
Some users had little tolerance for misunderstandings and unsatisfactory results. \textit{``I simply don't require these interactions. I'm not concerned about how it comprehends my requirements. My priority is to see results directly and as quickly as possible.''} $P_{4}$ added, \textit{``In fact, since communicating with AI is so complex, why not handle it myself?''} Usually, they knew exactly how to modify the design, and found it frustrating to translate their ideas into detailed descriptions. $P_{1}$ expressed: \textit{``OptiMuse was always trying to figure out what I meant by adding details or making descriptions more operated. I was kind of dizzy seeing the sentences became longer and longer...''} Additionally, $P_{9}$ reported a lack of professional vocabulary, which affected users' willingness to communicate about graphic design in the dialogue. Some interesting \textbf{interaction patterns} were noticed in these users:
\begin{itemize}[nosep]
\setlength{\leftskip}{-14pt}
    \item\textbf{Avoiding chats:} These users strongly opposed engaging in conversations with OptiMuse and devised various methods to circumvent dialogue. 
    Some of them added \textit{``Let's Have a try!''} after their commands to prompt OptiMuse to generate outcomes without engaging in conversations during the study. Once they discovered this trick, they employed it consistently.
    Another tactic involved typing ahead and anticipating responses. Some users never changed what they typed based on the actual replies. These users were actually iterating ideas by their own writing because they paid no attention to OptiMuse's textual feedback.
    \item\textbf{Selecting the best:} These goal-oriented users held clear expectations when selecting among choices. They chose the one closest to their expectations and iterated upon it to progressively approach a satisfactory result.
\end{itemize}

\subsubsection{Characteristics of Requirements in Design Practices}
Although participants' attitudes varied a lot towards OptiMuse, we observed consistency in the characteristics of requirements (requirement ambiguity, changes, and conflicts) among users using the same system. In the interview, participants agreed that the characteristics of requirements were unconsciously influenced by the difference between the two collaboration frameworks.

\textbf{Copilot: Strive for clarity, embrace change, and prevent conflicts.}
As we mentioned in Section \ref{sec:formativestudy}, ambiguity is almost unavoidable in requirement descriptions.
However, we observed how participants grapple with the challenge of being specific when using Copilot, as evidenced in their dialogues.
Additionally, during interviews, they self-reported the difficulty of being as detailed as possible to ensure Copilot's understanding. $P_{1}$ mentioned, \textit{``I was uncertain about how much detail I should provide for Copilot to ensure accurate understanding.''}
The requirements underwent slight changes to achieve a suitable degree or shift the focus to another aspect of the design. The background was initially instructed to be \textit{``darker,''} followed by \textit{``a little bit darker,''} then \textit{``still not dark enough,''} and finally, \textit{``well, try to make it a little lighter.''}
Similar iterations of descriptions were frequently employed in dialogues between humans and Copilot.
Conflicts of the requirement could not be well fixed in Copilot.
First, conflicts between asynchronous requirements like \textit{``darker and brighter''} were regarded as AI's lack of ability to produce appropriate outcomes.
Second, conflicts between the user and AI, namely, AI didn't give outcomes addressing the user's commands, were regarded as annoying mistakes. \textit{``I aimed for white titles, but for some reason, it consistently incorporated a gradient, which I found displeasing.``}, $P_{10}$ stated.
Conflicts within the design itself, the most tricky ones, were usually avoided in discussions. $P_{3}$ explained, \textit{``When I requested Copilot to justify both sides of all body text, the result was quite poor. I usually adjust the text box width when aligning the text. But I didn't bother explaining this to the AI.``}

\textbf{OptiMuse: Ambiguity sparks alternative ideas, changes and conflicts are the crucibles of creativity.}
There was a deliberate use of intentionally ambiguous descriptions observed in the interactions between participants and OptiMuse, as indicated by their dialogues.
Furthermore, participants expressed their willingness to take OptiMuse's opinions into account, reducing their concern about the precision of descriptions. $P_{3}$ said,\textit{``I was curious about OptiMuse's take on this design, so I provided a vague description of my ideas and awaited its response.''}
OptiMuse was designed to highlight changes in requirements and conflicts, which is often where participants reported gaining novel design ideas. \textit{``I was surprised when OptiMuse questioned my commands, but its suggestions appeared reasonable. So I decided to give it a try,''} $P_{8}$ added, \textit{``At least one of the selections turned out to be satisfactory. To be honest, I have never thought of doing so.''}
OptiMuse drove users to change requirements.
In OptiMuse, users' textual descriptions allow rapid exploration of the design space~\cite{kulkarni2023word}.
OptiMuse's suggestions (transitioning from dark mode to white mode, switching theme colors, rearranging each text module based on different rules, etc) increased designers' willingness to make bold changes. Users were prompted to decide among the changes, which encouraged reflection on each option.
Users took greater tolerance towards conflicts with OptiMuse.
Attitudes towards asynchronous conflicts were comparable in Copilot and OptiMuse. Nonetheless, such conflicts occurred less frequently in OptiMuse, as users were more focused on exploring new options rather than fine-tuning details.
Conflicts between AI and users' commands still happened in visual outcomes, but rather than simply dismissing them as mistakes, participants attempted to contemplate the merits of selective choices. \textit{``It felt like having an opinionated colleague. You know, negotiating and presenting options. So, I suppose there must be reasons behind whatever OptiMuse produces.''}, $P_{5}$ stated. Some participants were also receptive to OptiMuse's alternative suggestions during the dialogue.
\begin{table}[b]
\centering
\caption{Expected roles of AI agents and corresponding tasks.}
\setlength{\tabcolsep}{1.5pt}
\begin{tabular}{c c c c c c}
\hline
Roles & \begin{tabular}[c]{@{}c@{}}\thead{ Requirement\\Understanding}\end{tabular} & \begin{tabular}[c]{@{}c@{}}\thead{Requirement \\Analysis}\end{tabular} & \begin{tabular}[c]{@{}c@{}}\thead{Solution\\Exploration}\end{tabular} & Review & Execution \\ \hline
Executor                  &                                                                     &                                                                &                                                                &        & +         \\
Optimizer                 & +                                                                   &                                                                & +                                                              &        & +         \\
Colleague                 & +                                                                   & +                                                              & +                                                              & +      & +         \\
Expert                    & +                                                                   & +                                                              & +                                                              & +      & ?         \\ \hline
\end{tabular}
\label{tab:roles_tasks}
\end{table}
~\begin{table}[b]
\centering
\caption{Expected roles and adaptations within the human-AI co-design framework for different user groups.}
\setlength{\tabcolsep}{1pt}
\begin{tabular}{c c c c}
\hline
Roles     & Input                & Interaction             & Output               \\ \hline
Executor  & \small design               & \small confirm, revise         & \small optimized design     \\ 
Optimizer & \small design, requirement & \small confirm, revise         & \small optimized design     \\ 
Colleague & \small design, requirement & \small dialogue, select, remix & \small selective choices     \\ 
Expert    & \small design, requirement & \small dialogue, select, remix & \small selective choices     \\ \hline
\end{tabular}
\label{tab:roles_adaptation}
\end{table}

\subsection{Lessons We Learnt from Study II}
What altered people's preferences towards OptiMuse and Copilot? To answer this problem, we summarized people's different expectations of AI agents in human-AI co-design and their preferences.
Specifically, we try to address two questions inspired by Li et al.~\cite{li2023ai}: (i)What tasks are anticipated from AI agents, and (ii)What roles are expected of AI agents?

\subsubsection{What tasks are anticipated from AI agents?}
We adjusted the sequence of these two questions because tasks will be frequently discussed in formulating the definition of AI agents' roles.
The tasks are as follows:
\textit{Requirement Understanding:} comprehend users' design requirements conveyed through textual input;
\textit{Requirement Analysis:} engage in a dialogue with users to establish consensus regarding design objectives or operations. In other words, provide textual feedback towards users' textual input;
\textit{Solution Exploration:} present alternative design options in either textual or visual formats;
\textit{Review:} evaluate designs based on design knowledge, making comparisons with existing solutions;
\textit{Execution:} transform operational requirements into visual outputs.

\subsubsection{What roles are expected of AI agents?}
\label{sec:roles}
    Based on the interviews, our users expressed varying preferences and requirements regarding OptiMuse and Copilot. Specifically, they held different perceptions of the roles AI should play in graphic design. We categorized these roles as \textbf{expert}, \textbf{colleague}, \textbf{optimizer}, and \textbf{executor}, noting a diminished emphasis on creativity.
    \begin{itemize}[nosep]
    \setlength{\leftskip}{-22pt}
        \item[] \textbf{Expert: Give me an authoritative review of these options.} Out of the 1/12 participants, a minority expressed a preference for AI to function as an expert. An expert is tasked with providing reliable guidance on the existing design or design choices, as well as offering insights into alternative solutions. While it is essential for AI to grasp the designer's high-level requirements, the users place greater importance on AI's insights due to the inherent limitations of individual design skills. Requirement analysis plays a pivotal role, with a specific emphasis on elucidating AI's recommendations. This includes listing design principles, identifying issues, and drawing comparisons with exemplars. Ultimately, execution is just the icing on the cake, as what the target users seek is a reliable source of inspiration.
        \item[] \textbf{Colleague: Work with me as a different me.} Out of the 6/12 participants, a substantial portion preferred AI to function as a peer. More precisely, they envisioned an AI colleague on equal footing with the designer. A colleague AI is anticipated to comprehend user requirements to the fullest extent, showcasing this understanding through dialogue and thoughtful selections. In this context, AI's insights hold equal significance to the users, thereby enhancing the individual's design exploration and reflection. The outcome of execution remains essential, as end users seek not only potential creativity but also efficiency in saving time.
        \item[] \textbf{Optimizer: Whatever you do, make my design look better.} Out of the 2/12 participants, a few preferred AI to function as an optimizer. In this role, AI comprehends the initial design input and autonomously generates improved design alternatives. Users who lean towards this approach tend to be less inclined to review the original design or provide explicit optimization instructions. Instead, they repeatedly select the best options from a set of choices until they achieve a satisfactory solution or run out of patience.
        \item[] \textbf{Executor: Do just as I tell you.} Among the 3/12 participants, the prevalent choice was for AI to work only as an executor. An executor AI is expected to faithfully carry out users' commands. This specific type of user has clear goals and the possible operations required to achieve them, although they remain uncertain about the outcome until they actually put a hand to it. As a result, they depend on AI to reduce the manual effort involved in turning operations into visible results. They prefer to avoid requirement analysis, finding it cumbersome to explain operations to AI. Furthermore, they have no inclination to make selective choices because they wish to avoid the additional time and effort required for comparisons and selections among choices.
    \end{itemize}

    The tasks and adaptations within human-AI co-creativity systems corresponding to the expected roles of AI agents are listed in Table ~\ref{tab:roles_tasks} and Table ~\ref{tab:roles_adaptation}.

\section{Discussion}
\label{sec:discussion}
\subsection{Nonlinear Design for Co-creative Systems}
In popular tools like Copilot~\cite{copilot} and Midjourney~\cite{midjourney}, AI agents necessitate precise instructions and do not facilitate the remixing of various options (Fig.~\ref{fig:pipeline}).
They prioritize addressing the constraints inherent in the human-human co-design process (\textbf{DR4, DR5}).
As described in Kulkarni et al.'s work \cite{kulkarni2023word}, designers often refine their prompts multiple times and regenerate images until they achieve relatively satisfactory results.
Following this, they may choose to adopt either a portion or the entire outcome, either through manual editing or by blending prompts to generate new designs.
Such turn-based linear workflow hinders the natural nonlinear navigation among creative design stages~\cite{hatchuel_new_2003,pugh_total_1991, designcouncil,gero1990design,gero2004situated,howard_describing_2008}, which involves drawing inspiration from various parallel iterations and potentially modifying initial ideas along the way~\cite{Girotto2016collective}.
As a result, existing AI-assisted tools fail to consider the design requirements (Table. ~\ref{tab:DR}) because of their requirements for precise commands (fails to fulfill \textbf{DR1-2}) and restricted operations on outcomes (\textbf{DR3}).

Existing frameworks~\cite{negrete2014apprentice,kantosalo2016modes,guzdial2019interaction} on human-AI collaboration have shifted attention from AI as mere tools to more collaborative and creative agents.
However, they simply describe the co-design process as ``turns'' or ``iteration''.
We bridge this gap by formulating a nonlinear human-AI co-design framework featuring flexible communication and selective choices.
Following the nonlinear framework, OptiMuse enables parallel iteration~\cite{Girotto2016collective} and remixing~\cite{tan2022conflict,Cheliotis2014remix,Oehlberg2015remix,yue2019improving}.
Similar to Rezwana's finding in Penpal~\cite{rezwana2022understanding}, we observe enhanced user engagement in OptiMuse, where AI was perceived as more collaborative than a mere tool.
In summary, our framework serves as a compelling example of integrating theories (creative design process, human-AI framework) and practices (collective creativity strategies, Study I), which offers practical guidelines for the development of future human-AI co-creative systems.

\subsection{Roles of AI Agents in Co-creativity}
Desired roles of AI in human-AI collaboration have been widely discussed from different perspectives (e.g. categories of human-computer interaction for creativity~\cite{lubart2005can}, tasks in ideation~\cite{maher2012computational}, computational creativity~\cite{negrete2014apprentice}, work division in data storytelling~\cite{li2023ai,li2023far}).
While Lubart et al.~\cite{lubart2005can} stressed the diversity of users for AI roles, the characteristics of target users are under-discovered.

\begin{figure}[h]
    \centering
    \includegraphics[width=0.5\linewidth]{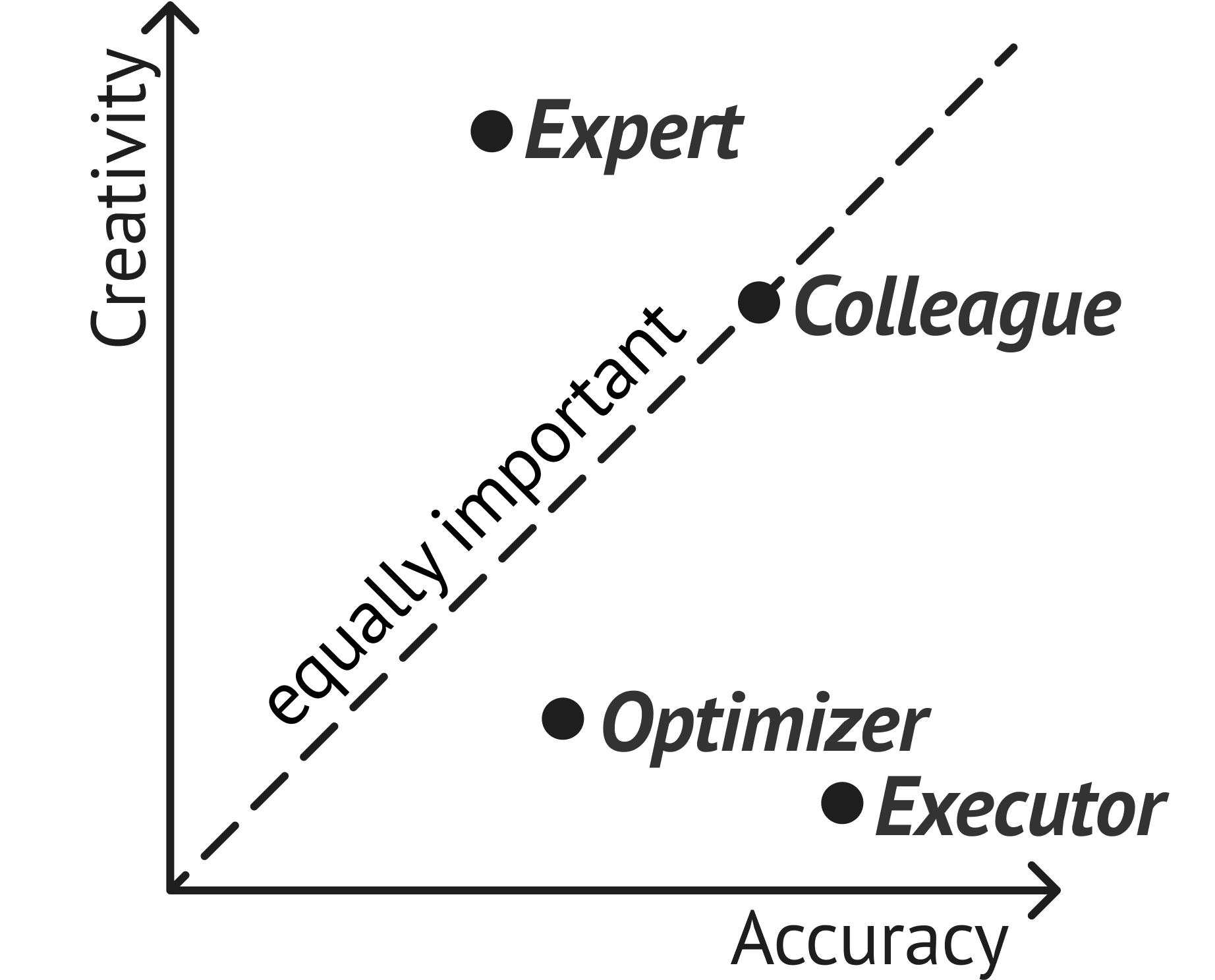}
    \caption{Participants' expectations regarding the creativity and accuracy required for different roles of AI agents.
    }
    \label{fig:woz_roles}
\end{figure}

Categorizing users' expected AI collaborators (Sec. ~\ref{sec:roles}), we propose four distinct roles for AI agents.
We mainly consider dimensions, namely creativity and accuracy, because both are the primary attributes that participants value in AI-assisted graphic design (Fig.~\ref{fig:woz_roles}).
Creativity refers to generating unconventional design ideas, while accuracy pertains to fulfilling design expectations.
Fig.~\ref{fig:woz_roles} illustrates the estimated levels of creativity and accuracy for each of the four AI roles, along with a guideline that deems both creativity and accuracy to be equally important.
When AI agents serve as \textit{experts}, designers with ill-defined tasks expect the highest level of creativity and demonstrate a relevant tolerance toward accuracy, in search of professional suggestions.
However, participants' expectations regarding the creativity of AI colleagues vary, but on average, they demand less creativity compared to AI experts.
When it comes to AI colleagues, users prioritize diverse perspectives over novelty.
Accuracy is also deemed equally important as effective communication between users and AI colleagues is crucial.
Conversely, AI optimizers and executors demand less creativity as users prioritize satisfaction over surprise.
The AI optimizer allows for a moderate level of accuracy tolerance, falling between the levels expected from AI experts and AI colleagues, since it does not involve input requirements.
The role of the executor necessitates the highest level of accuracy in executing users' commands, leaving little room for creativity.
In a human-AI co-creativity system, these AI roles may need to be implemented simultaneously or shifted due to the diverse needs of users.
There is no one-size-fits-all user, but rather a range of user profiles that can be further categorized for clarity and customization purposes.
Given the diversity and consistency of design challenges, users' intentions are multifaceted and may evolve throughout the creative design process.
Consequently, the development of a specific system must be meticulously planned in alignment with the intended user base and creative design scenarios.
For example, Study II highlights a divergent perspective towards OptiMuse and Copilot, indicating that the roles of AI should be accompanied by a redesign of user interfaces and AI actions in communication and selective choice should be adjusted accordingly.
    
\subsection{Limitations and Future Work}

One limitation of our research is the scope of user studies, which focused solely on graphic design as a creative design scenario and recruited participants with expertise in this field.
This restricted user sample may have resulted in incomplete findings.
Therefore, we view our guidelines for the nonlinear human-AI co-design framework as a starting point to enhance human-AI collaboration by emulating human collaborations.
We aim to inspire future research in various domains, such as creative writing, music composition, drawing, coding, and design ideation.

Another limitation of our study is the use of the WoZ methodology, which resulted in unreasonably long response times of approximately two minutes for each textual response.
This had a negative impact on usability and discouraged participants from engaging in co-design with AI agents.
However, it is worth noting that none of the participants expressed skepticism about our AI agent, and the identical delay in both systems allows for comparable System Usability Scale results.
Therefore, we view this study as a practical compromise to validate the nonlinear framework.
As AI technology continues to advance, we look forward to future research that will build upon and refine our findings by implementing AI-driven prototypes in accordance with our guidelines.

Admittedly, the importance of natural multimodal communication in human-human collaboration cannot be ignored.
When it comes to communication methods, both research~\cite{pan_human-computer_2023,shi2023Understanding}, design practices, and our formative studies have observed designers employing various forms, including text, voice, gestures, sketches, graphical operations, and even eye movements, to convey and exchange design ideas and reflections.
We acknowledge that the time-consuming nature of text-based interaction hindered the efficiency and user experience of human-AI collaboration, making it less comparable to human-human collaboration.
However, as highlighted by Ashktorab et al.~\cite{Ashktorab2020human}, social perceptions exhibit significant differences when considering interaction with an AI compared to interacting with a human~\cite{Ashktorab2020human}.
To summarize design requirements without falling into stereotypes of AI agents, we determined to gain insights directly from human-human collaboration; and to make it comparable to the existing AI tools, we chose text-based communication as the primary mode in our POC.
Shi et al.~\cite{shi2023Understanding} suggest that multimodal interaction in human-AI co-design is still in its early stages.
Looking ahead, we anticipate that ongoing research on multimodal interaction will lead to more flexible communication with AI systems.

\section{Conclusion}
\label{sec:conclusion}
This research focuses on the collaboration between humans and AI in creative design. 
Based on a formative study, we have established design requirements and guidelines that can inspire the development of nonlinear AI-assisted tools.
These guidelines aim to incorporate nonlinear characteristics into human-AI collaboration and address the limitations of human-human collaboration.
Following these guidelines, we propose a novel human-AI co-design framework and develop a POC, OptiMuse.
To evaluate our framework, we conduct a comparative study and implement a baseline tool that follows a linear human-AI workflow.
The findings demonstrate that OptiMuse significantly improves users' task completion rates and provides valuable insights for future AI-powered tools.

\begin{acks}
The work was supported by NSF of China (U22A2032, 62302440) and the Collaborative Innovation Center of Artificial Intelligence by MOE and Zhejiang Provincial Government.
\end{acks}
\bibliographystyle{ACM-Reference-Format}
\bibliography{main}

\appendix

\section{Participants Feedback in Study I}
\label{app:feedback}
\subsection{The Inherent Nonlinearity of Requirements.}
\begin{itemize}[nosep]
    \setlength{\leftskip}{-22pt}
    \item[] {\bf Requirement Ambiguity}
   Participants expressed their grievances drawing upon previous experiences:
    \textit{``When a client doesn't have any design know-how, it's hard for them to pinpoint the problem. All they can do is share their gut feelings about it.''} -B1
    
    \textit{``Some clients just throw out vague requests like, `Make this part look better.' or `Turn the background into a rainbow-infused black.' They never really break it down into operational instructions.''} - B4
    
    Some participants elucidated this viewpoint based on their own collaborative experiences as requirement proposers: 
   \textit{``Sometimes I have an idea of what I want to achieve, but I'm not sure how to visually express it because of limited design skills. So I can only describe it to the best of my ability.''} - A3
   
   \textit{``Sometimes I'm not sure what the problem is, and I can only say it doesn't look good.''} - B3
    \item[] {\bf Requirement Changes}
    For example, in A62, the designer meant to emphasize important parts but submitted ''change the color scheme'' as his initial modification requirement.
   \textit{``As we talked and exchanged ideas, my thoughts on making changes became clearer. I also realized new things that needed to be modified and discovered some issues with how I explained my ideas to the other person.''}- A6

    \item[] {\bf Requirement Conflict} 
        For instance, A13 stated, \textit{``It's kind of ugly to have text on a flag, but it's kind of boring without the background.''} Likewise, the contradiction between the amount of text and a neat layout is considered a common yet maddening situation.
\end{itemize}

\subsection{Pros and Cons for the Nonlinear Human-human Co-design Process.}
The co-design process is an imaginative voyage of discovery. 
% \No{4.1}
the presence of another designer brings forth a multitude of distinctions and divergences in many aspects.
\begin{itemize}[nosep]
    \setlength{\leftskip}{-22pt}
    \item[+]{\bf Inspiration to design in alternative ways.} 
        \textit{``Our (A1 and B1) collaboration worked wonders because we brought different viewpoints to the table, which complemented each other perfectly.''} said A1.\\
        \textit{``I never really thought about blending texts with the legend. It just didn't occur to me that the diagrams and the text should be connected. Then B2's suggestion hit me: why not change the legend from Chinese to English? That turned out to be a brilliant way to make the connection clearer.''} - A2\\
        \textit{``When I work alone, I tend to get stuck in my ways and miss out on learning new methods of presenting information. Take the `timeline' format in the experiment as an instance. While it may seem straightforward, I never considered placing the three phases on a timeline. Collaborating with others opened my eyes to this `innovative' approach, which served as a valuable knowledge enhancement.''}-B4\\
    \item[+]{\bf Collaboration accelerates the design progress under timely communication.} 
    \textit{``Communication can speed up the modification process, possibly calling off a faulty understanding of requirements or an unsatisfactory modification program upfront.``} - B3\\
    \textit{``It reduces tangling. When designing alone, I'm always struggling with whether it's going to work or not. But if there's another person to discuss, he can guide me, like, `Let's line up the worst piece first, and then we'll go back to struggling with the details of it.' ''} - A4
    \item[+]{\bf Misunderstandings are exposed and fixed swiftly through Q\&A.} 
    \textit{``The communication process helps to clarify your own modification needs, which are clearer and may be subject to change''} - B6\\
    \textit{``It might require more than a couple of questions to get what I'm trying to say.''} - A6\\
    \textit{``Misunderstandings sometimes inspire new ideas for modifications. For example, just now when he said one color for this module and another color for the other. I interpreted it as one color for each row because I hadn't thought about color matching module by module. But in fact, both of these modification options are quite good.''} - B6\\
    \item[-]{\bf Real-time outcomes lacking in human-human collaboration.} 
    \textit{``I want a `what you think is what you see' collaboration tool. Right now it's just me writing and sketching, but the draft is separate from the design itself, which doesn't change.''} - A4\\
    \textit{``I come up with modifications, but they don't always work out well in practice. You know, maybe you simply could not put all the words here, or in fact, those colors do not match.''} - B3
\end{itemize}

\begin{table*}[t]
\centering
\caption{Standard Operations for OptiMuse and Copilot}
\begin{tabular}{p{7cm}p{10cm}}
\toprule
\multicolumn{1}{c}{\textbf{User}} &
  \multicolumn{1}{c}{\textbf{OptiMuse}} \\ \midrule
\textit{Let’s have a try!} &
  Replace Image ABC. \\
A/B/C + comments &
  Update the system's preview image to A/B/C \\
\textit{NO} &
  \textit{I'm sorry if my modifications didn't meet your expectations. Please provide further information of your modification requirements.} \\
\textit{THX OptiMuse} &
  \textit{You're welcome! I'm happy to assist you. Best of luck with your design work!} \\
Requirements beyond the scope of design modifications &
  \textit{I apologize, but I can only answer questions related to design modifications.} \\
Requirements beyond the scope of the task &
  \textit{I'm sorry, your modification request exceeds the scope of this task.} \\ \midrule
\multicolumn{1}{c}{\textbf{User}} &
  \multicolumn{1}{c}{\textbf{Copilot}} \\ \midrule
Input valid texts &
  Replace Image \\
\textit{YES} &
  Update the system's preview image to the option \\
\textit{NO} &
  \textit{I'm sorry if my modifications didn't meet your expectations. Please provide further information of your modification requirements.} \\
\textit{THX OptiMuse} &
  \textit{You're welcome! I'm happy to assist you. Best of luck with your design work!} \\
Requirements beyond the scope of design modifications &
  \textit{I apologize, but I can only answer questions related to design modifications.} \\
Requirements beyond the scope of the task &
  \textit{I'm sorry, your modification request exceeds the scope of this task.} \\ \bottomrule
\end{tabular}
\label{tab:operations}
\end{table*}

\section{Instructions for the Wizard}
\label{app:wizard}
\textbf{Experimental Materials.} Figma project; Design*2*Task*2; Possible modification versions.\\
\textbf{Actions.} Chat in Figma; Export images from PowerPoint; Replace images in Figma; Change the Preview in Figma; Edit PowerPoint\\
\textbf{Imitating AI Features.}
    \begin{enumerate}[nosep]
    \setlength{\leftskip}{-12pt}
        \item Use punctuation at the end of sentences.
        \item Address the user as ``\textit{Nin}'' (formal).
        \item Maintain politeness by using phrases like ``\textit{thank you}'', ``\textit{I'm sorry}'', and acknowledging the user's correctness.
        \item Use consistent and appropriate language when referring to elements in the PPT (e.g., graphics).
        \item Occasionally rephrase the user's statements.
    \end{enumerate}
\textbf{Conversation Guidance}
    \begin{enumerate}[nosep]
    \setlength{\leftskip}{-12pt}
        \item Provide suggestions beyond the user's specified modifications to inspire them. For example, if the user wants to change the color of the text, you can suggest adjusting the background to complement the text color.
        \item Use your own judgment and offer proactive modification recommendations.
        \item When the user doesn't provide a clear modification request, guide them to be more specific, including specifying the modification area (e.g., overall, a specific image, a particular word, background, a certain color) and defining the modification scope (e.g., brightness, contrast, color temperature, hue, a specific color, a particular mood; font weight, font size, position, alignment, line spacing, letter spacing).
        \item When encountering vague requests, prompt the user to clarify their requirements, including specifying the modification area (overall, image, background, text) and the modification scope (brightness, contrast, color temperature, hue, specific color, particular feeling; font weight, font size, position, alignment, line spacing, letter spacing).
        \item Address conflicts in requirements: (i) When the user presents conflicting requirements at different times. Op: ``\textit{Your current request appears to contradict your earlier description of 'xxxx'.}'' (ii) When the user's request conflicts with suggestions from the AI assistant. Op: ``\textit{Are you sure about...?}'' (iii) When the modification request itself contains conflicts. Op: ``\textit{Please note that by increasing the font size without changing the text content, the text may overflow the text box.}''
        \item When the user changes their requirements, remind them of the changes. Op: ``\textit{Previously, you requested a darker background color. Now, you're asking for a warmer background color. Would you like to make the background color warmer and darker?}''
    \end{enumerate}
\end{document}